\documentclass{workshop}
\usepackage{amssymb}

\begin{document}

\title{
Heavy elements form short and long gamma-ray bursts
}

\author{
Daniel M. Siegel$^{1,2}$\\[12pt]  
%
$^1$  Perimeter Institute for Theoretical Physics, Waterloo, Ontario, Canada, N2L 2Y5 \\
$^2$  Department of Physics, University of Guelph, Guelph, Ontario, Canada, N1G 2W1\\
%
\textit{E-mail: dsiegel@pitp.ca} 
}

\abst{
The gravitational-wave detectors LIGO and Virgo together with their electromagnetic partner facilities have transformed the modus operandi in which we seek information about the Universe. The first ever-observed neutron-star merger---GW170817---confirmed the association of short gamma-ray bursts with neutron-star mergers and the production of heavy (r-process) elements. Based on recent theoretical and observational developments, I briefly present and discuss a conjecture, namely that compact accretion disks in both short and long gamma-ray bursts synthesize most of the heavy r-process elements in the Universe. The upcoming era of multi-messenger astronomy may allow us to verify or falsify this conjecture.
}

\kword{short GRBs --- long GRBs --- kilonovae --- r-process --- neutron-star mergers --- collapsars}

\maketitle
\thispagestyle{empty}

\section{Introduction}
\label{sec:intro}

It is a somewhat surprising fact that roughly 150 years after the periodic table of the elements has been established the cosmic origin of a large fraction of the elements listed in there still remains an open question. These rapid neutron capture (r-process) elements encompass nuclei of mass numbers $A\gtrsim 69$, just beyond the iron-peak elements, down to the lanthanides and actinides at the bottom of the periodic table---roughly half of all elements heavier than iron. They are created by capture of free neutrons onto seed particles (such as iron nuclei) under conditions in which the typical neutron-capture timescale is much shorter than the beta-decay timescale of the resulting nucleus.

The seminal papers by \citet{burbidge_synthesis_1957}, \citet{cameron_nuclear_1957}, and \citet{cameron_origin_1957} speculated that the very high neutron fluxes needed to trigger rapid neutron capture would require extreme astrophysical conditions. What was long thought to provide such neutron-rich environments---regular core-collapse supernovae---is now  disfavored as a production site of r-process elements. Theoretical results show that the high neutrino irradiation from the proto-neutron star formed as a result of the collapse resets the composition of outflows and leads to conditions unfavorable for the r-process \citep{qian_nucleosynthesis_1996,thompson_physics_2001,martinez-pinedo_charged-current_2012,roberts_medium_2012}. Observationally, core-collapse supernovae are disfavored based on rates and yields to synthesize r-process isotopes such as $^{244}$Pu in the solar neighborhood \citep{wallner_abundance_2015,hotokezaka_short-lived_2015}.

In this short proceedings contribution, I shall briefly state and discuss the following conjecture which points to both short and long gamma-ray bursts (GRBs) as the origin of r-process elements:\\

\noindent\textbf{Conjecture:} \emph{Outflows from compact (neutrino-cooled) accretion disks synthesize most of the heavy r-process elements in the Universe.}\\

Here, we shall focus on `heavy' r-process elements---elements beyond the second r-process peak ($A\gtrsim 130$), i.e., lanthanides and actinides. Those can be reliably traced in stellar spectra through europium (almost exclusively produced by the r-process), which is important when connecting astrophysical production sites with abundance observations in our galaxy through chemical evolution models. Light r-process elements might be co-produced by other sites, such as MHD supernovae \citep{winteler_magnetorotationally_2012-1,mosta_r-process_2018,halevi_r-process_2018}.

This conjecture is based on recent theoretical and observational developments: it draws from studies of nucleosynthesis in outflows from compact accretion disks in neutron star (NS) mergers (e.g., \cite{siegel_three-dimensional_2017,siegel_three-dimensional_2018,fernandez_long-term_2019,christie_role_2019,miller_full_2019-1}) as well as in collapsars \citep{siegel_collapsars_2019,miller_full_2019}. Observationally, it draws from the first detection of a neutron star merger---GW170817---by Advanced LIGO and Virgo \citep{abbott_gw170817:_2017}, its associated short GRB \citep{abbott_gravitational_2017-1} and the kilonova \citep{abbott_multi-messenger_2017}, the thermal counterpart that provided strong evidence for the synthesis of r-process nuclei in the merger ejecta (e.g., \cite{metzger_kilonovae_2019}, \cite{siegel_gw170817_2019} for overviews). Additionally, the conjecture is based on simple chemical evolution models that connect astrophysical production sites with spectroscopic abundance observations of r-process elements in our galaxy \citep{siegel_collapsars_2019,siegel_gw170817_2019}.

In view of the above conjecture, Sec.~\ref{sec:NS_mergers} briefly summarizes r-process nucleosynthesis in short GRBs (NS mergers); Sec.~\ref{sec:collapsars} provides the corresponding overview for long GRBs (collapsars). Aspects of chemical evolution are outlined in Sec.~\ref{sec:chemical_evolution} and Sec.~\ref{sec:discussion} provides further points of discussion. Conclusions are presented in Sec.~\ref{sec:conclusion}

\section{R-process in short GRBs}
\label{sec:NS_mergers}

Over the last twenty years, numerical simulations of the merger and post-merger phase in NS--NS and NS--black-hole binaries have established several mechanisms by which neutron-rich material is ejected from these systems. These include dynamical ejecta of a tidal and shock-heated nature \citep{ruffert_coalescing_1997-1,rosswog_mass_1999-1,oechslin_relativistic_2007,hotokezaka_progenitor_2013}, neutrino-driven and magnetically driven winds from a (meta-) stable remnant \citep{dessart_neutrino_2009,siegel_magnetically_2014,ciolfi_general_2017}, and outflows from a post-merger accretion disk \citep{fernandez_delayed_2013,just_comprehensive_2015,siegel_three-dimensional_2017}, the details and relative importance of which depend on binary parameters and the unknown equation of state of nuclear matter at supranuclear densities. 

Somewhat surprisingly, what had traditionally been believed to constitute the dominant mass ejection mechanism---dynamical ejecta from the merger itself---was disfavored by the observed properties of the GW170817 kilonova.
While the inferred ejecta parameters (ejecta velocity, mass, and lanthanide mass fraction) for the early blue emission of $v_\mathrm{ej}\approx 0.2-0.3\,c$, $M_\mathrm{ej}\approx 1-2\times 10^{-2}\,M_\odot$, and $X_\mathrm{Lan}\lesssim 10^{-4}$ are only marginally inconsistent with shock-heated ejecta for equations of state with very small neutron star radius, the red emission peaking on a timescale of a week requires $v_\mathrm{ej}\approx 0.07-0.14\,c$, $M_\mathrm{ej}\approx 4-6\times 10^{-2}\,M_\odot$, and $X_\mathrm{Lan} = 0.01-0.1$ \citep{nicholl_electromagnetic_2017-1,cowperthwaite_electromagnetic_2017-1,chornock_electromagnetic_2017-1,villar_combined_2017-1,kasen_origin_2017-1,kasliwal_illuminating_2017,mccully_rapid_2017-1,smartt_kilonova_2017-1,troja_x-ray_2017,pian_spectroscopic_2017-1,arcavi_optical_2017-2,perego_at_2017,coughlin_constraints_2018},\footnote{See \citet{smartt_kilonova_2017-1}, \citet{tanaka_kilonova_2017}, \citet{waxman_constraints_2018}, \citet{kawaguchi_radiative_2018} for different approaches.} which is in strong tension with theoretical predictions for dynamical ejecta (see Fig.~1 in \cite{siegel_gw170817_2019}). These kilonova parameters come with significant systematic uncertainties, which, however, unlikely affect this conclusion \citep{siegel_gw170817_2019}. I also refer the reader to \citet{metzger_welcome_2017}, \citet{siegel_gw170817_2019}, \citet{metzger_kilonovae_2019}, and \citet{radice_dynamics_2020} for reviews on the interpretation of the GW170817 kilonova event and further details.

The combination of high ejecta mass, low ejecta velocity, and lanthanide-bearing outflow of the red kilonova emission in GW170817 is consistent with outflows from a massive post-merger accretion disk.\footnote{These properties had been predicted shortly before the GW170817 event occurred \citep{siegel_three-dimensional_2017}.} Recent simulations in general-relativistic magnetohydrodynamics (GRMHD) with weak interactions find that roughly 30--40\% of the initial disk mass are ejected in unbound outflows \citep{siegel_three-dimensional_2017,siegel_three-dimensional_2018,fernandez_long-term_2019,christie_role_2019}. The slow outflow speeds of $\lesssim\!0.1c$ are the result of slow MHD winds which are further accelerated by the nuclear binding energy release of roughly 8\,MeV per baryon (corresponding to $\sim\!0.1c$) as the flow decompresses and free nucleons recombine into seed particles for the r-process \citep{siegel_three-dimensional_2018}. The neutron-rich conditions with mean electron/proton fraction $Y_e < 0.25$ required for lanthanide production are provided by the inner part of the accretion disk, which feeds sufficiently neutron-rich material into the outflows to keep the overall mean electron fraction favorable for lanthanide production. This is thanks to the high densities in the inner accretion disk, which forces electrons to become degenerate and thus favors electron over positron capture ($e^- + p\rightarrow n + \nu_e$ vs. $e^+ + n \rightarrow p + \bar{\nu}_e$); mild electron degeneracy is maintained thanks to a self-regulation mechanism \citep{chen_neutrino-cooled_2007-1,siegel_three-dimensional_2017,siegel_three-dimensional_2018}.

Massive post-merger accretion disks whose outflows dominate other r-process ejecta sources are expected to be ubiquitous. Provided that NS--NS mergers follow the galactic double neutron star mass distribution and given current constraints on the equation of state, \citet{margalit_multi-messenger_2019} estimate that $\sim\!70-100\%$ of all NS--NS mergers would avoid prompt collapse into a black hole. Excluding prompt collapse, recent simulations employing different equations of state find typical disk masses of $\sim\!0.1\,M_\odot$ \citep{radice_binary_2018}, which translate into outflows consistent with GW170817 (see \cite{siegel_gw170817_2019} for details). The fraction of light vs. heavy r-process elements produced in disk outflows depends on the mass of the disk \citep{miller_full_2019-1} as well as on the lifetime of a NS remnant \citep{lippuner_signatures_2017}. Taking GW170817 as a typical merger (its total mass is representative of a typical galactic double neutron star system), most mergers may be dominated by sufficiently short remnant lifetimes, in which case NS--NS mergers are dominated by lanthanide-bearing disk ejecta.

The contribution of r-process ejecta from NS--BH systems may be subdominant with respect to NS--NS mergers given expected yields and current constraints on their rates (NS--NS merges: 110-3840 Gpc$^{-3}$yr$^{-1}$; NS--BH mergers: $<610$ Gpc$^{-3}$yr$^{-1}$; at 90\% confidence; \cite{abbott_gwtc-1_2019}). In particular, a potentially large fraction of NS--BH systems---those with mass ratios significantly different from unity---may not give rise to ejecta at all, unless the black hole is rapidly spinning \citep{foucart_black-holeneutron-star_2012}. It is worth pointing out, however, that near-equal-mass mergers lead to negligible dynamical ejecta and are dominated by outflows from massive accretion disks \citep{foucart_numerical_2019,kyutoku_possibility_2020}, whose outflows are expected to be dominantly lanthanide-bearing. Systems with high mass ratios and high black hole spin can give rise to comparable amounts of dynamical and disk ejecta (e.g., \cite{kawaguchi_models_2016,foucart_dynamical_2017-1}).

\section{R-process in long GRBs}
\label{sec:collapsars}

Theoretical \citep{macfadyen_collapsars:_1999} as well as observational evidence connects long GRBs with the hyper-energetic explosions of massive stars stripped of their hydrogen envelopes \citep{woosley_supernova_2006}. Numerous analytical and numerical studies  have investigated the dynamics of the collapse to a black hole surrounded by an accretion disk and accretion shock, as well as the production of a successful relativistic GRB jet and supernova explosion \citep{macfadyen_supernovae_2001,fujimoto_magnetohydrodynamic_2006,uzdensky_magnetar-driven_2007,morsony_temporal_2007-1,bucciantini_relativistic_2008,lazzati_x-ray_2008-1,kumar_mass_2008,nagakura_jet_2011,lindner_simulations_2012,lopez-camara_three-dimensional_2013,batta_cooling-induced_2014}. 

Recent 3D GRMHD simulations of collapsar accretion disks including weak interactions and approximate neutrino transport find that collapsar accretion disks can give rise to r-process outflows, with an estimated mass of $\sim\!\mathrm{few}\times 10^{-2}\,M_\odot$ to $\lesssim\!1\,M_\odot$ depending on the stellar progenitor model \citep{siegel_collapsars_2019}. Stellar material collapsing onto the accretion disk with roughly equal numbers of protons and neutrons ($Y_e\approx0.5$) once in the inner disk is forced into neutron-rich conditions ($Y_e\ll 0.5$) favorable for the r-process by an analogous mechanism as previously discovered in NS mergers (electron degeneracy coupled to a self-regulation mechanism). While at very early stages during the collapsar accretion process, strong neutrino self-irradiation of the disk may prevent the synthesis of lanthanide-bearing winds \citep{miller_full_2019}, most of the r-process outflow is typically expected to be produced at intermediate accretion rates $\sim\!10^{-3}\,M_\odot\,\mathrm{s}^{-1}$ to $<10^{-1}\,M_\odot\,\mathrm{s}^{-1}$ \citep{siegel_collapsars_2019}, which give rise to lanthanide-bearing material.

The contribution of galactic r-process material from collapsars \emph{relative} to NS mergers can be estimated purely empirically by using the isotropic-equivalent energies (tracking the accreted mass) and rates of short versus long GRBs. This suggests that collapsars dominate the NS merger contribution by at least a factor of a few \citep{siegel_collapsars_2019}, owing to their larger isotropic-equivalent energies (larger accreted masses) which overcompensate for their lower rate of occurrence. This estimate is consistent with independent arguments based on chemical evolution models that take into account both NS mergers and collapsars (Sec.~\ref{sec:chemical_evolution}). Assuming that collapsars follow the star formation history and are responsible for most of the galactic r-process material, an empirical \emph{absolute} estimate of the collapsar per event yield can be obtained, which amounts to $\sim\!\mathrm{few}\times 10^{-2}-\mathrm{few}\times 10^{-1}\,M_\odot$ for standard assumptions \citep{siegel_collapsars_2019}. Calibration to the ejecta mass of GW170817 (assumed typical), this absolute estimate is consistent with the previous relative estimate. Finally, the typical per-event yield estimated theoretically based on simulations and progenitor models (see above) is consistent with the absolute empirical estimate. There are, admittedly, large uncertainties---nevertheless, it is encouraging that all independent estimates agree within predicted ranges; future studies will be able to shrink these uncertainties and test whether consistency prevails.

\section{Chemical evolution}
\label{sec:chemical_evolution}

Assuming GW170817 and its ejecta mass is typical of mergers, such systems could in principle account for most or all of the galactic r-process content. Notwithstanding the fact that such arguments assume a `closed box' and do not account for galactic outflows, there are a number of problems that afflict NS mergers as the dominant contributor to the galactic r-process when analyzing the synthesis of r-process elements over galactic history. I refer the reader to \citet{siegel_gw170817_2019} and references cited therein for an overview and more details on some of the issues. These include: prompt enrichment at low metallicities in halo stars, enrichment of ultra-faint dwarf galaxies (Reticulum II, Tucana III) as well as globular clusters (M15), and the assembly of r-process and alpha-elements in the galactic disk. Essentially, these issues result from the fact that NS binaries typically have significant systemic kick velocities (imparted by supernova explosions) and occur with significant delays with respect to star formation (due to the inspiral time).

In contrast to NS mergers, collapsars trace star formation history without significant delay ($\sim$few Myr due to their lifetime) and occur directly in the star-forming region that gave birth to the massive progenitor star. Prompt enrichment without spatial dislocation is favorable for r-process enrichment of halo stars \citep{van_de_voort_neutron_2019}\footnote{but see also \citet{macias_stringent_2018-1} for constraints in closed `single-supernovae' environments regarding the issue of iron co-production}, as well as for enrichment in ultra-faint dwarfs and globular clusters \citep{bonetti_neutron_2019,zevin_can_2019}. Furthermore, prompt enrichment by a dominant contribution of collapsars also reproduces the late-time trends of r-process elements relative to iron and alpha-elements in Galactic disk stars. Calibrating to these high-metallicity abundance observations, simple chemical evolution models taking into account both NS mergers and collapsars find that a dominant contribution of r-process elements from collapsars ($\gtrsim\!80\%$) is required to reproduce the observed trends \citep{siegel_collapsars_2019,siegel_gw170817_2019}, which is consistent with other independent estimates (see Sec.~\ref{sec:collapsars}).

\section{Discussion}
\label{sec:discussion}

It is worth pointing out some caveats and other points of discussion regarding the conjecture formulated in Sec.~\ref{sec:intro}
\begin{enumerate}
	\item If the NS--NS binary population does not follow the galactic double neutron star distribution, and instead is not dominated by near equal-mass systems, tidal ejecta can become a significant contributor to lanthanide-bearing ejecta besides disk winds (e.g., \cite{kiuchi_revisiting_2019}).
	
	\item A high-mass NS--NS binary population beyond the galactic distribution, which the recent GW190425 event---if indeed a NS--NS merger---may suggest \citep{the_ligo_scientific_collaboration_gw190425_2020}, could give rise to significant lanthanide-bearing tidal ejecta if such systems are dominated by strongly unequal mass systems \citep{dietrich_binary_2015}. However, if such a population is dominated by near equal mass systems, their contribution to r-process material would be minute, given that they would mostly undergo prompt collapse and thus leave little tidal and disk ejecta. Rates of GW170817-like events would currently not be affected by an additional population \citep{the_ligo_scientific_collaboration_gw190425_2020}.

	\item Realistic nucleosynthesis simulations of post-merger and collapsar accretion disks are still in its infancy. Sensitivity to weak interactions and a more detailed neutrino transport, more realistic initial conditions, and a larger part of the parameter space need to be explored to complement existing results. Additionally, more detailed chemical evolution studies are required to connect to abundance observations in various environments.

	\item The contribution of other sources to r-process elements, such as MHD supernovae, need to be further investigated. Stable jets in MHD supernovae that lead to lanthanide production likely require unrealistically high progenitor magnetic field strengths with the current treatment of neutrino interactions \citep{mosta_r-process_2018}. Given current rates, if MHD supernovae did produce significant amounts of lanthanides, their contribution would likely be minor, even if collapsars are neglected \citep{siegel_gw170817_2019}.

	\item Direct observational evidence for r-process nucleosynthesis only exists for NS mergers to date. R-process ejecta in collapsars is expected to lead to a near-infrared excess and features in the supernovae lightcurve and spectra at late times, once the surrounding supernova ejecta becomes transparent to emission from its r-process `core' (a `kilonova in a supernova'; \cite{siegel_collapsars_2019}). This might be observable for nearby long GRB events ($z\lesssim0.1$).
\end{enumerate}

\section{Conclusion}
\label{sec:conclusion}

In this proceedings contribution, I have briefly stated and discussed the conjecture that outflows from compact accretion disks in short and long GRBs synthesize most of the comic heavy r-process content. One exciting aspect of this conjecture is that it is at least to some extend observationally testable. 

Future multi-messenger observations of NS mergers will sample the NS binary population and provide more accurate estimates of rates and r-process yields as well as probe ejection mechanisms. Multi-messenger observations of NS--BH mergers will reveal their contribution to r-process nucleosynthesis and lead to a better understanding of the diversity of abundance distributions mergers can give rise to. This can be used to address open questions regarding diversity of observed abundance patterns for light r-process elements in halo stars and dwarf galaxies, as well as to address fundamental questions about nuclear properties and the universality of the strong r-process \citep{siegel_gw170817_2019}. Observations with future sensitive optical and near-infrared telescopes, such as the Large Synoptic Survey Telescope and the James Webb Space Telescope, may identify signatures of heavy r-process nuclei in GRB supernovae. Together with improved theoretical modeling, the upcoming era of multi-messenger astronomy may tell us whether accretion disks are indeed the preferred way nature provides the universe with r-process elements. This has important implications for astrophysics, nuclear physics, as well as chemical evolution and galaxy formation.\\


DMS acknowledges the support of the Natural Sciences and Engineering Research Council of Canada (NSERC). Research at Perimeter Institute is supported in part by the Government of Canada through the Department of Innovation, Science and Economic Development Canada and by the Province of Ontario through the Ministry of Colleges and Universities.

\bibliographystyle{aasjournal}

\bibliography{references.bib}



\label{last}

\end{document}